\documentclass[12 pt]{article} %latex2e
\usepackage{graphics}
\usepackage{amsmath,amssymb}
\usepackage{graphicx}
\usepackage{caption}
\usepackage{subcaption}
\usepackage{booktabs}
\usepackage[colorlinks=true, allcolors=blue]{hyperref}

\newcommand\br{\begin{eqnarray}}
\newcommand\er{\end{eqnarray}}
\newcommand\brs{\begin{eqnarray*}}
\newcommand\ers{\end{eqnarray*}}
\newcommand\be{\begin{equation}}
\newcommand\ee{\end{equation}}

\usepackage{fancyheadings}
\makeindex

\renewcommand{\sectionmark}[1]%
        {\markright{\thesection\ #1}}
\title{\Large Generalised reversible transformations and the inhomogeneous nonlinear Schr\"odinger equation
hierarchy }
\author{Sudipta Nandy\thanks{sudipta.nandy@cottonuniversity.ac.in} \ and \ Abhijit Barthakur \\ 
Cotton University,  Panbazar, Guwahati -781001, India}
\date{\today}
\begin{document}
\maketitle
\begin{abstract}
Under investigation is the nonlinear Schr\"odinger equation hierarchies and the 
reversible transformations. We propose a generalized reversible transformation between 
the  the generalized NLSE hierarchy  with focussing and defocussing nonlinearity and the   
NLSE hierarchy forced with  a linear potential term. The corresponding extended concept 
of classical dark and bright solitons of the forced hierarchy, accelerating due to  linear 
potential as well as due to the dispersion are obtained directly without resolving the 
nonisospectral  inverse scattering problem. We have identified a set of new constraints among 
the dispersion and the nonlinear coefficients  in the inhomogeneous NLSE hierarchy, 
which are preserved after the transformations. The reversible transformations allow us to 
encompass inhomogeneous NLS, HNLS and higher order equations belonging to the class 
of  nonisospectral  family of inverse  scattering problems to the  isospectral NLS class of 
equations and study them under a general mathematical framework. We hope that our 
analysis provides a mathematical platform to study inhomogeneous NLSEs as well as  
open up the possibility of new applications in physics.

\end{abstract}
{\small 05.45.Yv,  42.81Dp, 42.65.Tg} 

\section*{Introduction}\label{INTROD}
Possibility of optical soliton in fiber and in bulk medium was first proposed by Hasegawa and 
Tappert \cite{HASEGAWA73}. Two of the most discussed optical solitons are, spatial 
soliton, that do not spread spatially due to diffraction  and  
temporal soliton, that do not spread temporally  due to dispersion\cite{{ZS72},{JH88},
{MATVEEV91}}. The dynamics of  both spatial soliton and temporal soliton are 
described by the same  nonlinear Schr\"odinger  equation (NLSE) except that the spatial 
parameters are replaced by their temporal counterparts.   The 
applications of NLSE have spread into many  other branches of sciences over the 
past few decades. Reports of new applications  of NLSE  soliton are coming in every month.
NLSE soliton are among  the most widely  studied concept in nonlinear optics. 
Temporal soliton, due to its remarkable dispersion-less property  is suitable for all  
optical communication system\cite{HASEGAWA2003}. On the other hand, in case of 
spatial soliton the variety of nonlinearity accessible is much broader. A detail account of the 
spatial soliton and various types of nonlinearity are found in \cite{{SEGEV98},
{KIVSHAR03},{BISWAS06}}. Langmuir wave soliton in plasma physics
 is another example, where
coherent nonlinear structures  associated with electron plasma can be described with 
the NLSE\cite{{CHEN76}, {ANTIPOV78},{ZHEN15}}.  Matter wave soliton in 
Bose-Einstein condensate is yet another example, where
the behaviour of  large number of particles at  near absolute zero  temperature,  known as 
Bose-Einstein condensate can be described by Gross-Pitaevski equation
\cite{{SERKIN04},{RAJU05}, {ATRE06}} which in $1-D$ is the NLSE. 
The list continues.\\

\noindent   
Drawing an analogy with the time independent Schr\"odinger equation, solitons of the integrable 
NLSE are  sometimes referred to as {\it autonomous} soliton, that is a soliton solution 
of NLSE in the absence of an external  potential.  
The NLSE in presence of external potential is an inhomogeneous nonlinear equation and 
is not integrable in general. With certain potential however, the equation is found to be 
integrable and admit soliton solution.  The soliton solution  in presence of such potential  
is  called {\it nonautonomous}. In 1976 Chen and Liu in \cite{CHEN76} introduced 
inhomogeneity in terms of  a "gravitational" like potential to the  integrable NLSE, 
to describe the behaviour of the Langmuir wave in a linearly inhomogeneous plasma. 
In the same communication Chen and Liu  also introduced a symmetry transformation 
called the  {\it Tappert transformation}, which transform  NLSE with linear 
potential to NLSE without the potential and vice versa. The transformation is very 
much similar to the 
transformation introduced by Husimi and Taniuti\cite{Taniuti53}  to transform 
Schro\"dinger equation with a linear potential  into  the free Schr\"odinger equation. 
It would be interesting to check whether such symmetries can also be extended to  
NLSE hierarchies.   
% Later on the transformations for  other  potentials, namely  harmonic potential, 
% polynomial potential, tanh potential etc was introduced \cite{}. 

The spectral problem associated with the NLSE is well known\cite{AKNS73}, where the eigen 
values are time independent(isospectral). The introduction of inhomogeneity in NLSE causes 
a major change in the spectral problem. In presence of inhomogeneity the eigen values  are no 
longer time independent and they vary linearly with time.  
Consequently one has to solve the nontrivial  nonisospectral  inverse scattering problem. 
However, with the Tappert transformation \cite{CHEN76}  the NLSE with linear potential 
can be transformed into NLSE without potential and vice versa. Consequently one may obtain 
the  soliton solution  of the nonisospectral integrable models  by the reversible  transformation  
from the  soliton solutions of corresponding  isospectral NLSE, without going through 
the rigorous procedure of solving the nonisospectral IST problem. 

The transformations are however,  nontrivial in nature and 
hence  are applied  to only a handful of   inhomogeneous problems. 
Recently in \cite{SERKIN18} authors considered a  
"gravitational" like potential to the  integrable higher order nonlinear Schr\"odinger 
equations (HNLSE) in order to take in to account: 
(1) the sliding filter method of the noise separation from  femto-second pulse train, 
where the peak  frequency of the  sliding  frequency filter is assumed to be a linear 
function of time variable;  (2) the Raman induced self scattering shift.  The resulting 
forced HNLSE though  exactly solvable but is a case of nontrivial nonisospectral  inverse  
scattering problem. They used  reversible gauge transformations and obtained the soliton 
solution from the corresponding soliton solution of  isospectral HNLSEs. 
It is well known that  NLSE shares many of the mathematical properties with its hierarchy,
infinite conserved quantities for instance,  is one of them. The symmetry 
transformations  for NLSE and HNLSE suggest that there might be a general symmetry 
transformation between force free NLSE  hierarchy and  forced NLSE hierarchy.
 The usefulness and potential of such  transformations are  unquestionable as 
they provide a  bridge between the more nontrivial nonisospectral integrable models and 
the isospectral integrable models. One obtains the solutions without solving the IST 
problem. \\
 
In this paper we explore a generalized reversible transformation between the  the generalized 
NLSE hierarchy  with focussing and defocussing nonlinearity and the  generalized 
NLSE hierarchy forced with  a linear potential.  
We start with the  transformation  between force free NLSE and forced NLSE 
in the following section, and then obtain the transformations for  the second equation  and the 
third equation of the  hierarchy and then 
propose  a general transformation relation between the equations of forced and force 
free hierarchies. We also present a  discussion about the   soliton properties, 
considering bright and dark soliton in the inhomogeneous regime. We conclude with the 
summary of our results and future possibilities in the subsequent section. 

%%%%%%%%%%%%%%%%%%%%%%%%
%%%%%%%%%%%%%%%%%%%%%%%%
%%%%%%%%%%%%%%%%%%%%%%%%

\section*{Reversible transformations} 

Consider the  inhomogeneous equation:
\be
\label{NLSE}
{\bf i} q_t + \sum_{i=1}^{\infty}   S_{2i}\{q\} +
 {\bf i}   P_{2i+1}\{q\}   - V(x,t) q  = 0 
\ee
where, $S_{2i}\{q\}$ and $ P_{2i+1}\{q\}$  are respectively the even and odd operators of the 
NLSE hierarchy. 
For instance, $S_2\{q\}= \frac{1}{2} D_{20} q_{xx} + \sigma R_{20} (|q|^2 q) $ is 
the NLSE operator.  The subscripts ($x$, $t$ ) denote the partial derivatives with respect to 
$x$ and $ t$ respectively.   
The inhomogeneous term namely, $V(x,t)= 2 \lambda(t)  x $ is a
gravitational like potential \cite{SERKIN2018Aug}. $\lambda(t)$ is the
time derivative of the spectral parameter . Note that the role of  $x$ and $t$ 
in NLSE are exchanged  when applied to optics. For the spectral parameter, linear in $t$,  
 $\lambda(t) = \lambda$, is a constant.  
In absence of  the inhomogeneous term the equation represents  the NLSE hierarchy generalized 
for focusing and defocusing nonlinearity.
In absence of the last term in eq. \ref{NLSE} the $S_{2i}$ and $P_{2i+1}$ are determined 
form the infinite conserved charges with appropriate Poisson bracket 
\cite{{HASEGAWA2003}, {CHAWDURY17}}.  It may be noted here that
NLSE hierarchy equations with $P_{2i+1}=0$  are different from the equations with  
$S_{2i}=0$  at least from one aspect, that is  in the second case the equations 
are the real modified KdV equation hierarchy. The distinction is expected to appear also in 
the soliton solutions of forced equations. 
With $S_{2}=  \frac{D_{20}}{2} q_{xx} +  \sigma R_{20}|q|^2 q $  and
$S_{2i}|_{i>1}=0 $, $ P_{2i+1}=0$, eq. \ref{NLSE} reduces to the  integrable 
inhomogeneous  NLSE \cite{CHEN76} for focusing or defocusing nonlinearity, 
according to the sign of  $\sigma$.
\be
\label{NLSE1}
{\bf i} q_t + \frac{D_{20}}{2} q_{xx} + \sigma R_{20} |q|^2 q  - 2 \lambda(t) q x =0
\ee
The  "$+$ sign" for $\sigma$ accounts for the focusing  nonlinearity and the  "$-$ sign" 
accounts for the defocusing nonlinearity. In optical applications $(x$, $t)$
represent the dimensionless time in a comoving frame and the normalized 
propagation distance respectively. The forced  eq. \ref{NLSE1} is a case of 
nonisospectral IST problem and is exactly solvable. An easier alternative approach 
to obtain the solution is to use  Tappert transformation\cite{CHEN76}:  
\br
\label{H1}
\begin{aligned}
E(T(x,t), Z)&=q(x,t) e^{i \phi(x,t)}\\
\phi(x,t)&= 2\lambda x t + \frac{2}{3} D_{20}\lambda^2 t^3  \\
T(x,t)&= x + D_{20} \lambda t^2  \\
Z&=t 
\end{aligned}
\er
and  convert eq. \ref{NLSE1} into free NLSE:
\be
i E_Z  + \frac{1}{2} D_{20} E_{TT} + \sigma R_{20} |E|^2 E =0
\ee
The well known NLSE soliton solution  may be obtained through 
several methods, namely IST method for isospectral problems \cite{AKNS73},
Hirota method \cite{JH88}, Ba\"cklund transformations \cite{MATVEEV91} etc. 
The solutions for focusing ($\sigma=+1$) and defocusing  ($\sigma= - 1$) nonlinearity 
respectively are: 
\br
\label{NSOL1}
E = \begin{cases} 2 \eta \sqrt{\frac{D_{20}}{ R_{20}}} 
sech (2 \eta T + 4 D_{20}\eta \kappa Z )
e^{(-i 2 \kappa T + i 2 D_{20} ( \eta^2 - \kappa^2) Z)}; &  bright \ soliton (\ \sigma = 1)\\
2 \eta  \sqrt{\frac{D_{20}}{ R_{20}}} tanh ( 2 \eta T +  4 D_{20} \eta \kappa Z  ) 
e^{(-i 2 \kappa T - i 2 D_{20} ( 2 \eta^2 + \kappa^2) Z)} ;  &    dark \ soliton (\ \sigma = - 1)
\end{cases}
\er
Finally  through reverse transformation:  
\br
\label{H1R}
\begin{aligned}
q(x(T,Z),t)&=E(T, Z ) e^{- i \psi(T, Z)}\\
\psi(T, Z) & = 2\lambda T\ Z - \frac{4}{3} D_{20} \lambda^2 Z^3  \\
x & = T  -  D_ {20}\lambda Z^2  \\
t&=Z
\end{aligned}
\er
the soliton solutions of   eq. \ref{NLSE1} are obtained. The solutions are: 
\begin{eqnarray}
\label{NSOL1}
q =   \begin{cases}  2 \eta  \sqrt{\frac{D_{20}}{ R_{20}}}
& sech (2 \eta x + 2 \eta D_{20} \lambda t^2  +4 D_{20}\eta \kappa t )
e^{-i 2 \kappa x  -i 2 \kappa D_{20} \lambda t^2 + i 2 D_{20} ( \eta^2 - \kappa^2) t}\\
~& \times e^{-i (2\lambda x t + \frac{2}{3} D_{20} \lambda^2 t^3)}; \quad
  bright \ soliton (\ \sigma = 1)\\
2 \eta  \sqrt{\frac{D_{20}}{ R_{20}}}
& tanh ( 2 \eta x + 2 \eta D_{20} \lambda t^2+  4 D_{20} \eta \kappa t  ) 
e^{-i 2 \kappa x -i 2 \kappa D_{20} \lambda t^2 - i 2 D_{20} ( 2 \eta^2 + \kappa^2) t} \\
~& \times e^{-i (2\lambda x t + \frac{2}{3} D_{20} \lambda^2 t^3)}; \quad
dark \ soliton (\ \sigma = -1)
\end{cases}
\end{eqnarray} 
It may be noted that  the reversible transformations(\ref{H1}, \ref{H1R}) do not depend on  
parameter $\sigma $. That is  Tappart transformation is independent of type of nonlinearity, 
focusing or defocusing. 
The inhomogeneity in the equation  introduces many changes in the soliton dynamics, such as in 
soliton's velocity and  phase, which are two of the principle quantities controlling 
the dynamics of soliton. 
For instance a    dark NLSE soliton with  group velocity "$V_g = 2 D_{20}\kappa $ " and 
phase velocity "$V_{ph} = \frac{1}{\kappa} D_{20} (2 \eta^2 + \kappa^2) $"  
are changed to $ 2 D_{20} \kappa +  D_{20} \lambda t  $ 
and  $ \frac{- D_{20}(2 \eta^2 +\kappa^2) + 2 D_{20} \kappa \lambda t 
+ D_{20} \lambda^2 t^2 + \lambda x }{\kappa + \lambda t} $  respectively
in the inhomogeneous regime.  The group velocity and the phase velocity are now linear function 
of 't', that is the motion of soliton is  "accelerated" in the presence of external potential. Further 
note that the soliton also changes it's colour during motion as it's frequency changes 
quadratically with time. \\

The same approach may also be adopted to higher order NLSE with the linear external potential.
Next let us consider  the second equation of the NLSE hierarchy, and rewrite the eq. \ref{NLSE} 
with
$S_{2}=  \frac{D_{20}}{2} q_{xx} +\sigma R_{20}|q|^2 q $  and 
$ P_{3}= D_{30} q_{xxx} + \sigma R_{30} 6 |q|^2 q_x  
+ \beta \sigma R_{30}  (|q|^2)_x q $,
\begin{equation}
\label{Rv1}
{\bf i} q_t +\frac{1}{2} D_{2 0}q_{xx} + \sigma R_{20} (|q|^2 q) 
+ {\bf i}( D_{30} q_{xxx} + \sigma R_{30}6 |q|^2 q_x  
+ \beta \sigma R_{30}  (|q|^2)_x q)  - 2 \lambda q x = 0
\end{equation}
We shall refer eq. \ref{Rv1}  as FHNLSE for future reference.  
In absence of  the inhomogeneous term   eq. \ref{Rv1} represents
three known cases of the hierarchy, namely the complex modified KdV equation(CMKdVE) 
($D_{20}= R_{20}=0$; $\beta=0 \ or \ 3$)\cite{SASA91};  Hirota 
equation(HE) ($\beta=0$) \cite{HIROTA73} and Sasa Satsuma equation(SSE) 
($\beta=3$)\cite{SASA91}  provided the parameters $D_{20}$,  $D_{30}$,  
$R_{20}$, $R_{30}$  satisfy the Hirota condition, namely 
 $ D_{20} R_{30} = D_{30} R_{20} $.     
Eq. \ref{Rv1} is important for the description of  ultrashort pulse propagation in fiber 
in presence of the effects such as, third order dispersion, self steepening 
and stimulated Raman Scattering in an inhomogeneous medium. The potential term, 
$V(x,t)= 2 \lambda(t)  x $  in the equation may  account for the sliding frequency 
method of the noise separation from a soliton \cite{{HASEGAWAKODAMA1995},
{BURTSEV1997}}.  Secondly, the 
inhomogeneous term also describe the change in the  frequency of the pulse due to 
Raman induced delayed nonlinear  response \cite{{SERKIN07},{GPA95}}.
The parameter $\sigma$ has the same significance as it has in the NLSE (eq. \ref{NLSE1}).   

The transformations of the  FHNLSE (eq. \ref{Rv1}) for $\sigma(= +1)$ is given 
in \cite{SERKIN18} for three different cases for HNLSE. Here we rewrite the
reversible  transformations in a generalised form, which would represent transformations 
for both focusing and defocusing cases  ($\sigma = \pm 1)$ and would transform
eq. \ref{Rv1} into the corresponding force free HNLSE. 
The transformations are:
\begin{eqnarray}
\begin{aligned}
E(T(x,t), Z)&=q(x,t) e^{i \phi(x,t)}\\
\phi(x,t)&= 2\lambda x t + \frac{2}{3}D_{20}\lambda^2 t^3 +2 D_{30}\lambda^3 t^4  \\
T(x,t)&= x +D_{20}\lambda t^2   +4 D_{30} \lambda^2 t^3  \\
Z&=t 
\end{aligned}
\end{eqnarray}
and the  converted force free equation is:
\begin{equation}
\label{FRv1}
{\bf i} E_Z +\frac{1}{2} D_2(Z) E_{TT} + \sigma R_2 (z)(|E|^2 E) 
+ {\bf i} ( D_{30} E_{TTT} + \sigma R_{30}  6 |E|^2 E_T  + 
\beta \sigma R_{30} (|E|^2)_T E) =0
\end{equation}
where, $D_2(Z) =  D_{20} + 12 D_{30}\lambda Z$ and 
$R_2(Z) = R_{20} + 12 R_{30}\lambda Z$. 
Eq. \ref{FRv1} corresponds to  free  
Hirota equation,  for  $\beta=0$; free Sasa-Satsuma equation, for $\beta=3$ and free 
CMKDV equation for  $ D_{20} = R_{20} = 0 $. Notice that  after the transformation the 
Hirota condition namely,  $D_2(Z) R_{30} = D_{30} R_2(Z) $  is still  satisfied.
At this point it may be worthwhile to mention that the force free Hirota equation and the 
Sasa Satsuma equation can be converted into CMKDV equation with  gauge 
transformations\cite{{HIROTA73},{SASA91}}.  Let us consider 
soliton solution of one of these cases, Hirota  equation for instance, 
\br
\label{Fsol}
E(T,Z) =   \begin{cases} 2 \eta \sqrt{\frac{D_{30}}{R_{30}}} 
                sech(2 \eta T  + 4 D_{20}\eta \kappa Z + 
                          + 8 \eta D_{30} (3 \kappa^2 - \eta^2 ) Z                     
                          +  24 D_{30} \eta \kappa \lambda Z^2 )\\
                \times  e^{ -2 i \kappa T +  2 i D_{20} ( \eta^2 - \kappa^2) Z
                 -   8 i \kappa D_{30} (\kappa^2 - 3 \eta^2 ) Z
                          +  12 i D_{30} (\eta^2 - \kappa^2  ) \lambda Z^2 }; \\
                \quad bright \  soliton \ ( \sigma= 1) \\
                2 \eta \sqrt{\frac{D_{30}}{R_{30}}} 
                 tanh(2 \eta T +  4 D_{20}\eta \kappa Z 
                         + 8 \eta D_{30}(3\kappa^2 + 2 \eta^2) Z 
                         + 24  D_{30} \eta  \kappa  \lambda Z^2 )\\
                 \times   e^{-i (2\kappa T  +2 D_{20} ( 2 \eta^2 + \kappa^2) Z
                         + 8  D_{30}  \kappa (\kappa^2 +  6\eta^2)Z         
                         +  12 D_{30}(\kappa^2 + 2 \eta^2)\lambda Z^2) 
                          } ; \\
                  \quad dark \  soliton  \ ( \sigma=- 1) 
                  \end{cases}
\er
Now with the following reverse transformations:
\begin{eqnarray}
\begin{aligned}
q(x(T,Z),t)& = E(T, Z) e^{i \psi(T,Z)}\\
\psi(T,Z)&  = - 2\lambda T Z + \frac{4}{3} D_{20} \lambda^2 Z^3 + 6 D_{30}\lambda^3 Z^4  \\
x(T,Z)& = T - D_{20} \lambda Z^2 -  4 D_{30} \lambda^2  Z^3  \\
t(Z) &= Z
\end{aligned}
\end{eqnarray}
the $1$-soliton solution  of the  forced HNLS equation form eq. \ref{Rv1} is obtained, 
\br
\label{Rvsol}
q(x,t) = \begin{cases}2 \eta \sqrt{\frac{D_{30}}{R_{30}}}
                                 sech(2 \eta x + 4 D_{20}\eta \kappa t
                                 + 8 \eta D_{30} (3 \kappa^2 - \eta^2 ) t  
                                 + 24 D_{30} \eta \kappa \lambda t^2 \\ 
                                 + 2  \eta D_{20} \lambda t^2
                                 + 8 \eta D_{30} \lambda^2 t^3 )                                    
                                  e^{-i( 2 \kappa x + 2 \kappa D_{20} \lambda t^2
                                         +  8 \kappa D_{30} \lambda^2 t^3 
                                         +  8 D_{30}\kappa ( \kappa^2 - 3 \eta^2 ) t )}\\
                                 \times  e^{-i(  -  2 D_{20} ( \eta^2 - \kappa^2) t  
                                     -  12 D_{30} (\eta^2 - \kappa^2  ) \lambda t^2 
                                    +  2 \lambda x t + \frac{2}{3} D_{20} \lambda^2  t^3 
                                    + 2 D_{30} \lambda^3  t^4)     }\\
                                       \quad  bright \  soliton (\sigma=1); \\
              2 \eta \sqrt{\frac{D_{30}}{R_{30}}} 
              tanh( 2\eta x + 4 D_{20}\eta \kappa t  
                + 8 \eta D_{30}  (3\kappa^2 +2 \eta^2) t  
                + 24  \eta D_{30}  \kappa \lambda t^2  \\
                + 2 \eta D_{20}  \lambda t^2 
                + 8 \eta D_{30} \lambda^2 t^3 )
                e^{-i(2\kappa x   +  2 \kappa D_{20} \lambda t^2  
                   + 8\kappa D_{30} \lambda^2 t^3                      
                   + 8 \eta D_{30}  \kappa (\kappa^2 +6\eta^2)t )}\\
             \times e^{-i( 2 D_{20} ( 2 \eta^2 + \kappa^2) t
                   + 12 D_{30} (\kappa^2 + 2 \eta^2)\lambda t^2                  
                   +  2 \lambda x t  + \frac{2}{3} D_{20} \lambda^2  t^3 
                   + 2 D_{30} \lambda^3  t^4     
               )} \\
              \quad dark \ soliton (\sigma = - 1);
\end{cases}
\er  
The soliton solutions pair (eq. \ref{Rvsol}) is the nonautonomous  soliton solution of 
eq. \ref{Rv1} and is interpreted as  a coloured (multi wavelength) soliton 
\cite{MAIMSTOVE99} because  its frequency is changing with '$t$'  during 
propagation. The group velocity of  nonautonomous NLS soliton  the nonautonomous  HNLS 
solitons are  fundamentally different in the sense that in the former case 
soliton's velocity  is reversed after reaching a minimum  and in the later case  
soliton's velocity increases in the same direction after reaching a minimum.   
Secondly the  third order dispersion(TOD) causes an asymmetric 
broadening of the soliton pulse that is the higher amplitude soliton moves faster, 
a phenomenon also exhibited by KdV soliton\cite{DAS89} which is because of the presence of 
TOD term in the equation. The consequence of dependence of the soliton velocity  on the 
amplitude is that the noise(having lower amplitudes) 
can be filtered out from the ultrashort soliton pulses. \\
 
The transformation discussed above may also be extended 
to  NLSE with higher order dispersion and nonlinear terms. In the literature study of  
such equations is rare, other than a few  reported theoretical  models 
\cite{{PORSEZIAN97},{YANG2018}, {DAI2018},{SERKIN2018Aug}}. 
However, recent development of  extremely short pulse (auttosecond), analysis of  
higher order terms  is gaining momentum again. Effects of the individual terms such as  
forth order dispersion (FOD) \cite{GPA95}, quintic  nonlinearity 
\cite{{AKHMEDIEV01},{BISWAS06},{KOMAROV05}} are studied separately
in different physical contexts. For instance, in  
dispersion flattened fibers, it is found that  FOD is 
significant\cite{GPA95} compared to TOD, such equation is not integrable but 
admit "autosoliton" solution. Recent  reports suggest that  $CDS_xSe_{1-x}$ doped glasses 
shows  fifth order susceptibility and  parabolic form of refractive index\cite{BISWAS03}.
NLSE in presence of either FOD or quintic nonlinearity is also not integrable but there exist 
solitary  waves solutions for these equations too.  Theoretical and experimental investigation on 
higher order derivatives of nonlinear terms are not available in the literature. This is due to 
the fact the detail  analysis on  the effects of these terms as well as the collective behaviour 
of  all the terms simultaneously  is extremely difficult task  at present and  require further 
advancement  of both  theory and experimental technology.  
However, to start with theoretically  it is always convenient to consider  an integrable  model, 
which contains  these terms. \\

Next in our analysis we consider  equations having quintic nonlinearity and fourth order 
dispersion term along with the inhomogeneous term.  That is along  with $S_{2}$ and  
$P_{3}$,  we also consider the term  $S_4 = D_{40} q_{xxxx} +  R_{40}
(6 |q|^4 q  +  2 \sigma q^2 q^*_{xx} +  4 \sigma |q_x|^2 q 
+6 \sigma q_x^2 q^* + 8 \sigma |q|^2 q_{xx}) 
- 2\lambda(t) q x $  and $ S_{2i+2}, \  P_{2i+1}=0$ for $ i \ge 2$, 
then  eq. \ref{NLSE} reduces to

\br
\begin{aligned}
\label{HRv1}
 {\bf i} q_t  + \frac{1}{2}D_{20} q_{xx}  + \sigma R_{20} |q|^2 q 
+ {\bf i}( D_{30} q_{xxx}  +  \sigma R_{30} |q|^2 q_x )
+  D_{40} q_{xxxx}   \\
+  R_{40}( 6 |q|^4 q  +  2 \sigma q^2 q^*_{xx} 
+  4 \sigma |q_x|^2 q + 6 \sigma q_x^2 q^* + 8 \sigma |q|^2 q_{xx}) 
-   2\lambda(t) q x = 0  
\end{aligned}
\er
With $\sigma=1$, eq.\ref{HRv1} with constant coefficients is the 
Lakshmanan-Porsezian-Deniel equation  \cite{{PORSEZIAN97}, {ANKIEWICZ14}, 
{ANKIEWICZ14A}},  attached with  the force term. In future we refer eq.\ref{HRv1}
as forced Lakshmanan-Porsezian-Deniel equation(FLPDE).  
Notice that if $D_{20}=D_{30}=R_{20}=R_{30}=0$, then  FLPDE
reduces to the following equation: 
\br
\begin{aligned}
\label{HRv2} 
 {\bf i} q_t & +  D_{40} q_{xxxx} \\
+ R_{40} ( 6 |q|^4 q  & +  2 \sigma q^2 q^*_{xx} 
+ 4 \sigma |q_x|^2 q + 6 \sigma q_x^2 q^*  +  8 \sigma |q|^2 q_{xx}) 
-  2\lambda(t) q x = 0  
\end{aligned}
\er
Note that this equation is also referred  as the Lakshmanan-Porsezian-Deniel equation
and is a hierarchy of the CMKdVE. 
We consider  additional two equations, which admit  soliton solution. First one is
FLPDE ( eq. \ref{HRv1}) in absence of the  NLS operator ($S_2$): 
\br
\begin{aligned}
\label{HRv3}
 {\bf i} q_t & +  
+ {\bf i}( D_{30} q_{xxx}  +  \sigma R_{30} |q|^2 q_x )
+  D_{40} q_{xxxx}   \\
+  R_{40}( 6 |q|^4 q  &+  2 \sigma q^2 q^*_{xx} 
+  4 \sigma |q_x|^2 q +6 \sigma q_x^2 q^* + 8 \sigma |q|^2 q_{xx}) 
 - 2\lambda(t) q x = 0  
\end{aligned}
\er
and the second one  is  FLPDE (eq. \ref{HRv1}) in absence of the Hirota operator ($P_3$): 
\br
\begin{aligned}
\label{HRv4}
 {\bf i} q_t &+ \frac{1}{2}D_{20} q_{xx}  + \sigma R_{20} |q|^2 q
+  D_{40} q_{xxxx}  \\
+  R_{40}( 6 |q|^4 q  &+  2 \sigma q^2 q^*_{xx} 
+  4 \sigma |q_x|^2 q +6 \sigma q_x^2 q^* + 8 \sigma |q|^2 q_{xx}) 
-   2\lambda(t) q x = 0  
\end{aligned}
\er
Note that the structures of  integrable hierarchies are highly rigid. Each of the hierarchies 
of the NLSE contains dispersion and nonlinear terms in  a fixed ratio. However, the 
nice mathematical symmetries among the hierarchy makes it convenient to study 
the complex higher order equations. \\

Let us introduce the following gauge  transformation  and 
transformation of variables: 
\begin{eqnarray}
\label{TR2}
\begin{split}
\begin{aligned}
E(x(T,Z),t)& = q(x, t) e^{i \phi(x,t)}\\
\phi(x,t)&  = 2\lambda x t + \frac{2}{3}D_{20}\lambda^2 t^3 
+2 D_{30}\lambda^3 t^4 - \frac{16}{5} D_{40}\lambda^4 t^5  \\
T(x,t)& =  x  +  D_{20}\lambda t^2   +  4 D_{30} \lambda^2 t^3 -   
8 D_{40}\lambda^3 t^4 \\
Z(t) &= t
\end{aligned}
\end{split}
\end{eqnarray}
which converts the forced eq. \ref{HRv1} into  the following 
a free  higher order NLSEs with variable coefficients, \\
\begin{eqnarray}
\label{HNLSTR}
\begin{aligned}
{\bf i} E_Z + D_{40} E_{TTTT} + \sigma R_{40} (6  |E|^4 E +   |E_T|^2 E
+   6 E_T^2 E^* + 2   E^2 E^*_{TT} + 8 |E|^2 E_{TT} ) \\
+ {\bf i} (D_3(Z) E_{TTT} + 6 \sigma R_3(Z)|E|^2 E ) + \frac{D_2(Z)}{2}  E_{TT} 
+ R_2(Z) \sigma |E|^2 E)= 0 
\end{aligned}
\end{eqnarray}
where 
$D_2(Z)= D_{20} + 12 D_{30} \lambda Z - 48 D_{40} \lambda^2 Z^2$, 
$ D_3(Z)= D_{30} - D_{40}8\lambda Z $,             
$R_2(Z)= R_{20} +12 D_{30} \lambda Z - D_{40} 48 \lambda^2 Z^2$, 
and  $R_3(Z)= R_{30} - D_{40}8\lambda Z$.\\

Note that the transformation eqs.\ref{TR2} with  $D_{20}= D_{30}=
=R_{20}= R_{30}=0$,
that is,
\begin{eqnarray}
\label{TR3}
\begin{split}
\begin{aligned}
E(x(T,Z),t)& = q(x, t) e^{i \phi(x,t)}\\
\phi(T,Z)&  =  2\lambda x t  - \frac{16}{5} D_{40}\lambda^4 t^5  \\
T(x,t)& = x   -  8 D_{40}\lambda^3 t^4 \\
Z(t) &=t
\end{aligned}
\end{split}
\end{eqnarray}
convert eq.  \ref{HRv2} into eq. \ref{HNLSTR} with
$D_2(Z)=  - D_{40}48 \lambda^2 Z^2$, 
$ D_3(Z) =  - D_{40} 8\lambda Z $, $R_2(Z)=  - 48 D_{40}\lambda^2 Z^2$, 
and  $R_3(Z)=  - D_{40} 8\lambda Z$.   

The transformations  from  eq. \ref{HRv3} to eq. \ref{HNLSTR} is accomplished
from eq. \ref{TR2} with $ D_{20}=0$. Consequently in the transformed equation,     
$D_2(Z)=   12 D_{30} \lambda Z -  48 D_{40} \lambda^2 Z^2$, 
$ D_3(Z)= D_{30} - 8 D_{40} \lambda Z $,             
$R_2(Z)=  12 R_{30} \lambda Z - 48 D_{40} \lambda^2 Z^2$, 
and  $R_3(Z)= R_{30} -  8 D_{40} \lambda Z$.

similarly  the transformations  from  eq. \ref{HRv4} to eq. \ref{HNLSTR} is accomplished
by using eqs. \ref{TR2} with $ D_{30}=0$. Consequently, 
$D_2(Z)= D_{20}   -  48 D_{40} \lambda^2 Z^2$, 
$ D_3(Z)=  - 8 D_{40} \lambda Z $,             
$R_2(Z)= R_{20}  - 48 R_{40} \lambda^2 Z^2$, and  
$R_3(Z)= - 8 R_{40} \lambda Z$.

Further note that similar to the Hirota constraint \cite{DIANOV85} in HNLS equations, 
we identify a set of new constraints among the dispersion and nonlinear coefficients,
namely, $D_{20} R_{40}= D_{40}  R_{20}$; $D_{30} R_{40}= D_{40} R_{30}$ and  
$D_{20} R_{30}= D_{30} R_{20}$  which are preserved under the transformations  
eqs. \ref{TR2}. Similar constraints are expected from the equations as one goes higher 
in the hierarchy. Number of constraints increases  as $~^nC_2$, where $n$ is the 
order of the equation in the NLSE hierarchy.\\

Let us consider $1$- soliton solutions of  the force free eq. \ref{HNLSTR} :
\br
\label{HHSOL}
E(T,Z) = \begin{cases} 
 2\eta \sqrt{\frac{D_4}{R_4}}
sech(2 \eta T  + 4 D_{20} \eta \kappa Z +  8 D_{30}\eta (3 \kappa^2 - \eta^2 ) Z 
                      + 24 D_{30} \eta \kappa  \lambda  Z^2 \\
                       -  64 D_{40} \eta \kappa (\kappa^2 - \eta^2) Z
                       -  32 D_{40} \eta ( 3 \kappa^2 - \eta^2 ) \lambda Z^2  
                       -  64 D_{40}\eta \kappa \lambda^2 Z^3 ) \\
\times e^{ - i 2 \kappa T + i 2 D_{20} ( \eta^2 - \kappa^2) Z  
                    -  i 8 D_{30}\kappa ( \kappa^2 - 3 \eta)^2 ) Z  
                    + i 12 D_{30}( \eta^2 - \kappa^2 ) \lambda Z^2 }\\
                  \times e^{   i 16 D_{40} ( \kappa^4 - 6 \kappa^2 \eta^2 + \eta^4) Z 
   -  i 32 D_{40} ( \eta^2 - \kappa^2 ) \lambda^2 Z^3 
   + i 32 D_{40} \kappa (\kappa^2 - 3 \eta^2)\lambda Z^2};  \\
\quad bright \  soliton \ ( \sigma = 1) \\
~\\
2 \eta \sqrt{\frac{{D_3}}{R_3}} 
tanh(
   2 \eta T  + 4 D_{20} \eta \kappa Z + 8 D_{30}\eta (3 \kappa^2 + 2 \eta^2 ) Z 
                  + 24  D_{30} \eta \kappa  \lambda  Z^2   \\   
                  - 64 D_{40} \kappa \eta \lambda^2 Z^3 
                  - 64 D_{40}\eta  \kappa ( \kappa^2 + 2 \eta^2) Z        
                  - 32 D_{40}\eta (3 \kappa^2 + 2 \eta^2 ) \lambda Z^2 )  \\
\times  e^{-  i 2 \kappa T - i 2 D_{20} ( 2 \eta^2  + \kappa^2) Z  
          - i 8 D_{30} \kappa (\kappa^2 + 6 \eta^2 ) Z  
           + i 12 D_{30} ( 2 \eta^2 + \kappa^2) \lambda Z^2}\\
        \times e^{  i 16 D_{40} (\kappa^4 + 12 \kappa^2 \eta^2 + 6 \eta^4) Z  
        + i 32 D_{40} ( 2 \eta^2 + \kappa^2) \lambda^2 Z^3  
        + i 32 D_{40}\kappa ( \kappa^2 + 6 \eta^2 ) \lambda Z^2
         }\\
 \quad dark \  soliton \ ( \sigma = - 1); 
\end{cases}
\er
we shall obtain the soliton solution of one of the solutions of  
eq. (\ref{HRv1}-\ref{HRv4}).  For instance, the soliton solution  of the first of these 
equations and the  reverse transformations for this equation is: 
\begin{eqnarray}
\begin{aligned}
q(x(T,Z),t) & = E(T, Z) e^{- i \phi(T, Z)}\\
\phi(T,Z) &  = 2\lambda T Z   - \frac{4}{3}D_{20}\lambda^2 Z^3  
                         -  4 D_{30} \lambda^3 Z^4
                        + \frac{64}{5} D_{40}\lambda^4 Z^5\\
x(T,Z))& = T  -  D_{20}\lambda Z^2   -  4 D_{30} \lambda^2 Z^3 +   
8 D_{40}\lambda^3 Z^4  \\
t(Z) & = Z
\end{aligned}
\end{eqnarray}
Thus the soliton solution of the forced eq. \ref{HRv1} is :
\br
\label{HHSOLTR}
q(x,t) = \begin{cases} 
2 \eta \sqrt{\frac{D_4}{R_4}}
sech(2 \eta x + 4 D_{20} \eta \kappa t + 2\eta D_{20} \lambda t^2   
        +  8 D_{30} \eta (3 \kappa^2 - \eta^2 ) t \\
         + 8\eta D_{30} \lambda^2 t^3 
          + 24 D_{30} \eta \kappa \lambda t^2            
           - 64 D_{40} \eta \kappa (\kappa^2 - \eta^2) t 
            - 64 D_{40} \eta \kappa \lambda^2 t^3 \\ 
             - 32 D_{40}\eta (3 \kappa^2 - \eta^2 )  \lambda t^2  
              - 16\eta D_{40}\lambda^3 t^4) 
 e^{- i 2 \kappa x  -  i 2 D_{20}\kappa \lambda t^2 
            + i 2  D_{20} (\eta^2 - \kappa^2) t }\\
\times e^{- i \frac{2}{3}D_{20}\lambda^2 t^3      
                 - i 8  D_{30} \kappa ( \kappa^2 - 3 \eta)^2 ) t
                  - 4 i  D_{30} \lambda^3 t^4
                   - 8 i D_{30} \kappa   \lambda^2 t^3
                    + i 16  D_{40} ( \kappa^4 - 6 \kappa^2 \eta^2 + \eta^4) t }\\
\times e^{ i 16  D_{40} \kappa \lambda^3 t^4 
                   - i 32  D_{40} ( \eta^2 - \kappa^2 ) \lambda^2 t^3 
                    + i 32   D_{40}\kappa (\kappa^2 - 3 \eta^2)\lambda t^2
                      + i \frac{16}{5} D_{40}\lambda^4 t^5
                        - i 2\lambda x t}; \\
\quad  bright \ soliton \ ( \sigma = 1) \\
~ \\
2 \eta \sqrt{\frac{D_4}{R_4}}     
tanh( 2 \eta  x  + 4 D_{20}\eta \kappa t + 2 D_{20} \eta  \lambda t^2 
        + 8 D_{30} \eta (3 \kappa^2 + 2 \eta^2 ) t  \\
         + 8 D_{30} \eta  \lambda^2 t^3   
          + 24 D_{30} \eta \kappa \lambda t^2 
            - 64 D_{40} \eta  \kappa ( \kappa^2 + 2 \eta^2) t  
             - 64 D_{40} \kappa \eta \lambda^2 t^3  \\ 
              - 32 D_{40} \eta (3 \kappa^2 + 2 \eta^2 ) \lambda t^2
               - 16 \eta D_{40}\lambda^3 t^4 )  \\
e^{ - i 2 \kappa x - i \frac{2}{3}D_{20}\lambda^2 t^3    
            - i 2  D_{20} ( 2 \eta^2 + \kappa^2) t }\\ 
\times e^{-  i 2 D_{20}\kappa \lambda t^2 
                  -  i 8  D_{30}\kappa (\kappa^2 + 6 \eta^2 ) t 
                  -  i 8 D_{30}  \kappa \lambda^2  t^3 
                    - 4 i  D_{30} \lambda^3 t^4
                      + i 16 D_{40} (\kappa^4 + 12 \kappa^2 \eta^2 + 6 \eta^4) t }\\
\times e^{i 32 D_{40} ( 2 \eta^2 + \kappa^2) \lambda^2 t^3 
                      + i 16  D_{40} \kappa \lambda^3 t^4
                       + i 32  D_{40} \kappa ( \kappa^2 + 6 \eta^2 ) \lambda t^2 
                         - i 2 \lambda x t + i \frac{16}{5} D_{40}\lambda^4 t^5}; \\ 
\quad dark \ soliton \ ( \sigma = -1);
\end{cases}
\er
Solution of the rest of the eqs. (\ref{HRv2}-\ref{HRv4}) may similarly  be obtained with the 
corresponding transformations discussed above.    
The analysis can be carried forward to even higher equations, such as  equations in 
\cite{{YANG2018}, {CHEN2018}} along with the potential  term. We notice a few observations in 
the above analysis.
First, solitons of odd order hierarchy have different space-time profile in general than the
soliton of even order hierarchy, an observation also noticed in \cite{CHAWDURY17} but for 
force free NLSE hierarchy.  In the odd order 
hierarchy soliton the $Sech$ argument contains  even order polynomial   in $\lambda t$ where as 
the even order hierarchy the the argument contains an odd  order polynomial   in $\lambda t$.  
Secondly, as $\lambda \rightarrow 0$ the nonautonomous solitons reduces to 
their autonomous counterparts.  
Thirdly each of the higher order equations   are reducible to 
the lower order equations by removing the higher order parameters, for instance,  
the  FLPDE  reduces to forced HNLSE  with  $D_{40}=R_{40}=0$ and the 
corresponding  nonautonomous solitons reduce to the  autonomous soliton of HNLSE. 
Similarly forced HNLSE reduces to the  forced NLSE with $D_{30}= R_{30}=0$. Above 
observations  show that our analysis is consistent with the earlier analysis, with 
force free NLSE hierarchy.
The simple form of reversible transformations suggests that the generalized 
reversible transformations between  the force free NLSE hierarchy and the 
forced NLSE hierarchy follow a general form: 
\be
E(T(x,t), Z(t))= q(x,t) e^{i\phi(x,t)}
\ee
Where  the phase $\phi (x,t)$ and the variable $T(x,t)$ are expressed as a series:
\br
\begin{aligned} 
\phi &=2 t (\lambda x+ a_0 (\lambda t)^2+ a_1 (\lambda t)^3 +.......);\\
T&= x + v(t)\ t ;\\
v(t)&= b_0 (\lambda t)+ b_1 (\lambda t)^2 +.......;\\
Z &=t 
\end{aligned}
\er
where $a_1$, $a_2$, $a_3$, ...., $b_0$, $b_1$, $b_2$, $\cdots$ 
are constants, to be determined.\\
For instance for forced  NLSE: \\
$a_0=\frac{1}{3}D_{20}$, 
$b_0= D_{20}$ and  $a_i,\ b_i =0 \ for \ i\ge 1$ ;\\
for forced CMKdVE:\\
$a_1=D_{30}$,  
$b_1= 4 D_{30}$ and $a_i,\ b_i =0 \ for \ i\ne 1$ ;\\
for forced Hirota equation and Sasa Satsuma equation: \\
$a_0=\frac{1}{3}D_{20}$, $a_1=D_{30}$, 
$b_0=D_{20}$, $b_1=4 D_{30} $, $a_i,\ b_i =0 \ for \ i\ge 2$ ;\\
for FLPDE without NLS and Hirota operator:\\
$a_2=\frac{-8}{5}D_{40}$,  $b_2= - 8 D_{40}$ and $a_i,\ b_i =0 \ for \ i\ne 2$ ;\\
forced HNLSE with cubic, quintic  nonlinearities and fourth order dispersion, 
FLPDE eq. (\ref{HRv1})\\
$a_0=\frac{1}{3}D_{20}$, $a_1=D_{30}$, $a_{20} =\frac{-8}{5}D_{40}$, 
$b_0=D_{20}$, $b_1=4 D_{30}$, $b_2= - 8 D_{40}$,  and $a_i,\ b_i =0 \ for \ i\ge 2$ ;\\
HNLSE with cubic, quintic nonlinearities and fifth order dispersion \\
$a_0=\frac{1}{3}D_{20}$, $a_1=D_{30}$, $a_2=\frac{-8}{5}D_{40}$, 
$a_3=\frac{-16}{5}D_{50}$
$b_0=D_{20}$, $b_1=4D_{30}$, $b_2= - 8 D_{40}$, $b_{30} = -16D_{50} $   
and $a_i,\ b_i =0 \ for \ i\ge 3$ ;\\
and so on. 
\begin{figure*}[htbp]
\begin{subfigure}{.5 \textwidth}
\centering
\includegraphics[width=1 \textwidth]{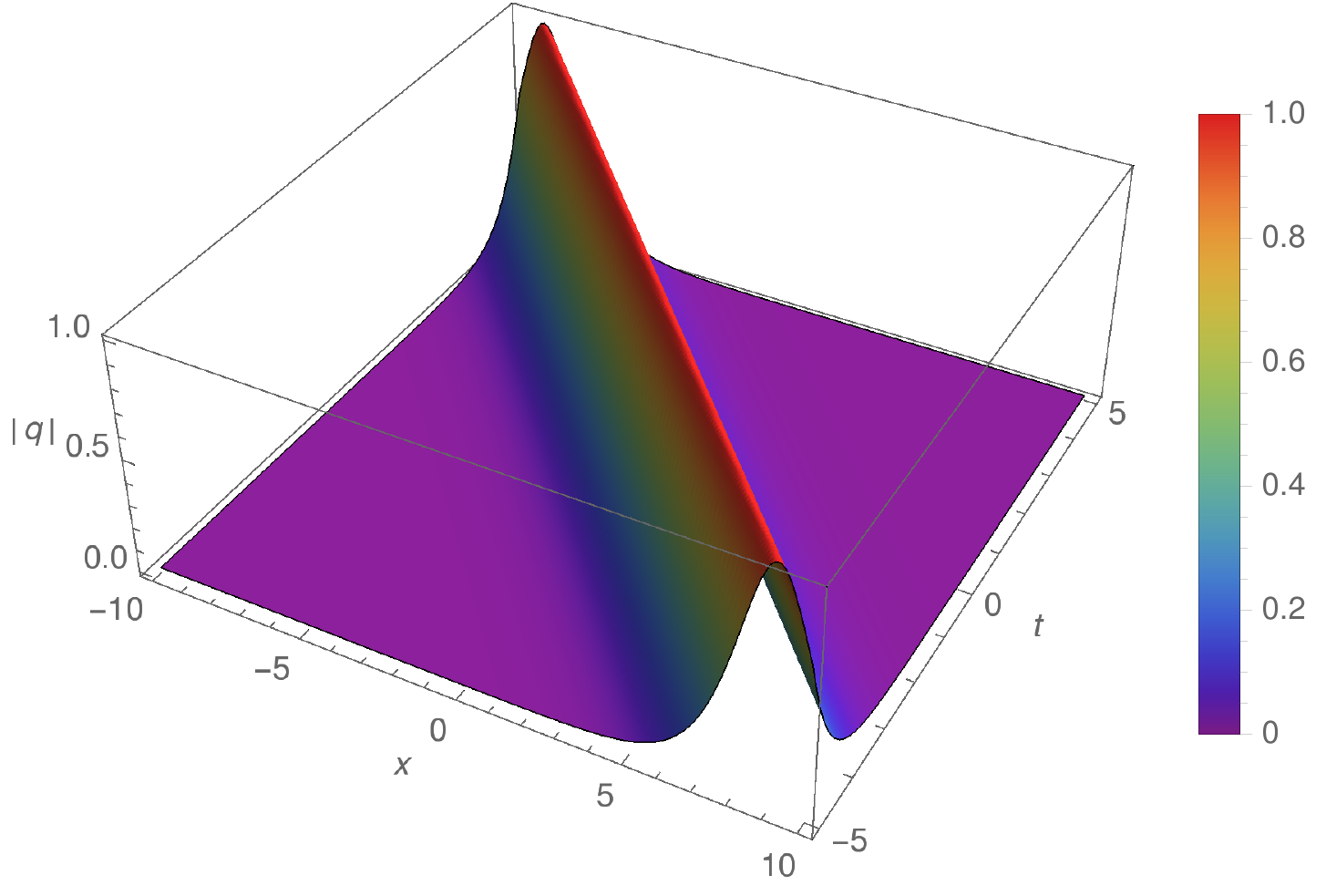}
\caption{}
\end{subfigure}%
\begin{subfigure}{.5 \textwidth}
\centering
\includegraphics[width=.9\textwidth]{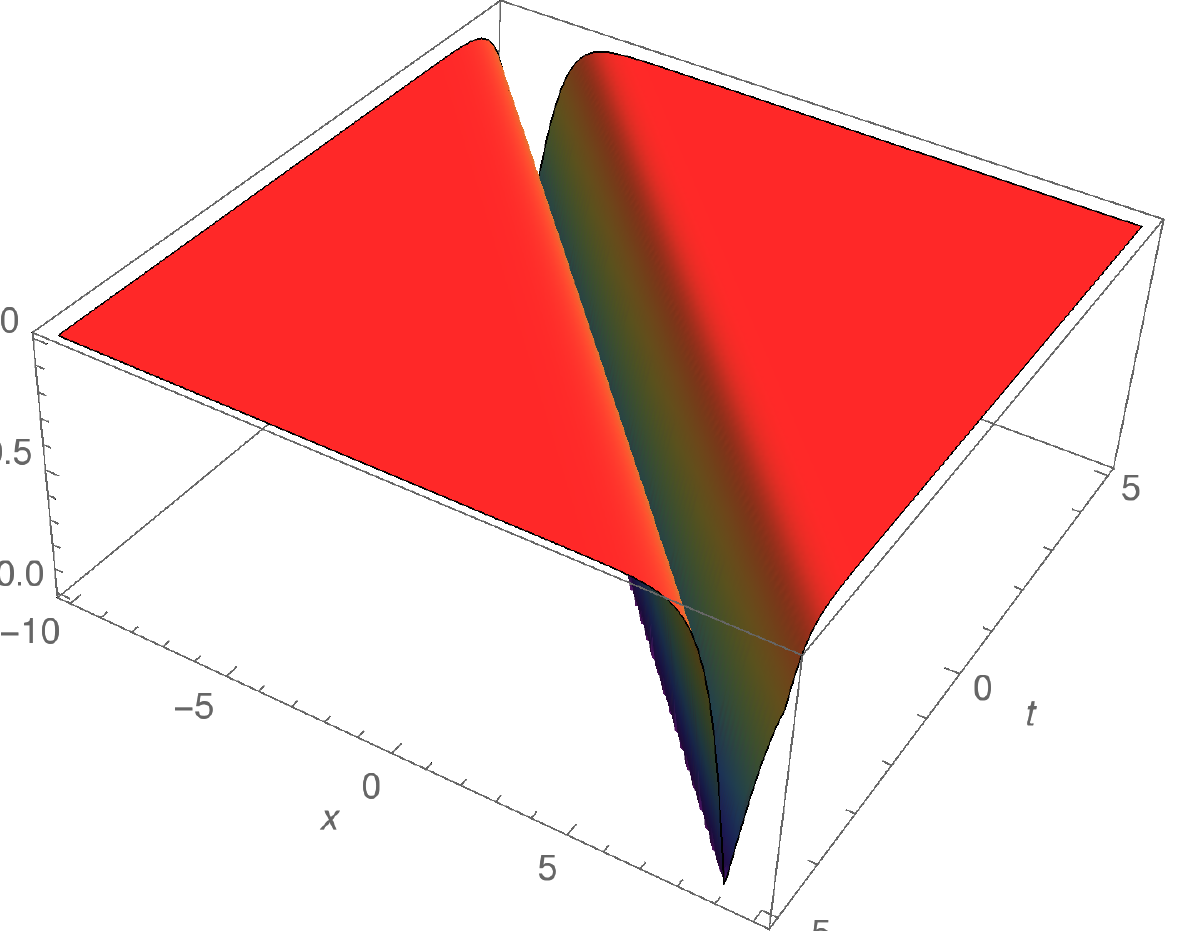}
\caption{}
\end{subfigure}%
\caption{Evolution of an autonomous NLSE bright(a) and dark(b) solitons at constant 
velocity with $D_{20}=R_{20}=1 $, $\lambda = 0 $ and soliton parameters 
$\eta=0.5$ and $\kappa= 0.9$}
\label{fig-R1}
\end{figure*}

\begin{figure*}[htbp]
\begin{subfigure}{.5 \textwidth}
\centering
\includegraphics[width=1 \textwidth]{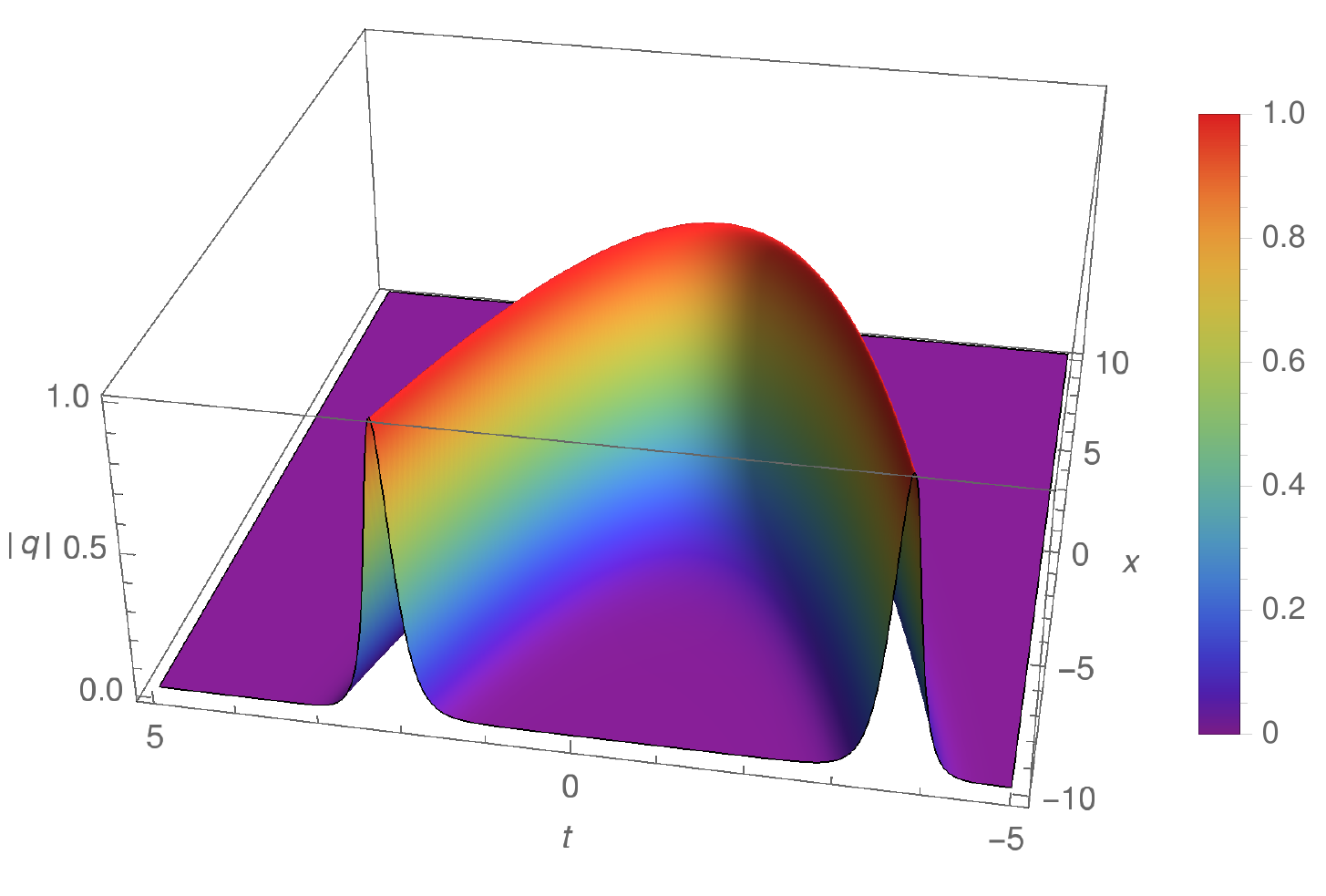}
\caption{}
\end{subfigure}%
\begin{subfigure}{.5 \textwidth}
\centering
\includegraphics[width=.9\textwidth]{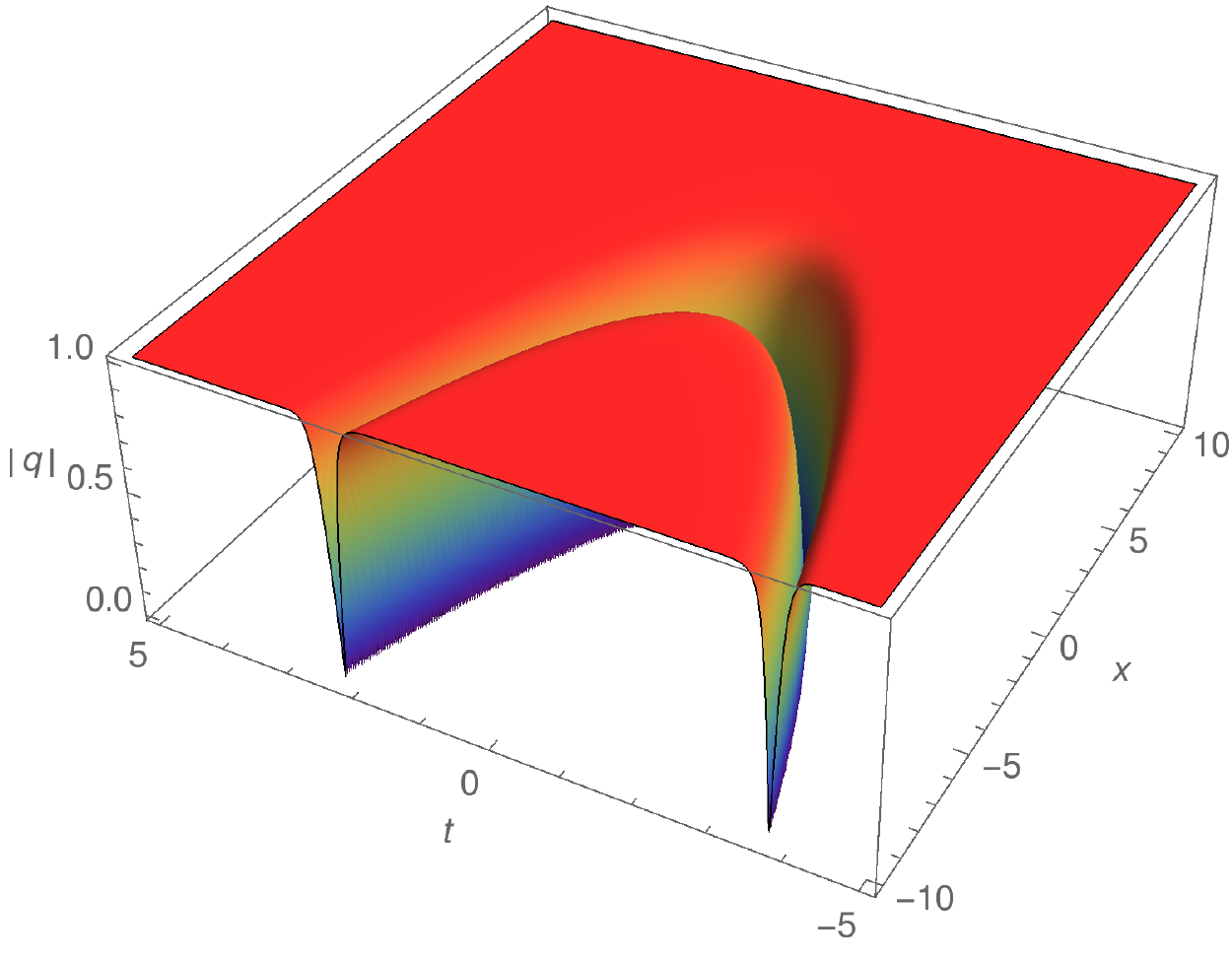}
\caption{}
\end{subfigure}%
\caption{Evolution of a nonautonomous NLSE bright(a) and dark(b) solitons with 
$D_{20}=R_{20}=1 $, $\lambda = 1.2$ and soliton parameters $\eta=0.5$ and $\kappa= 0.9$
 (\ref{Rvsol})}
\label{fig-R2}
\end{figure*}

\begin{figure*}[htbp]
\begin{subfigure}{.5\textwidth}
\centering
\includegraphics[width=1\textwidth]{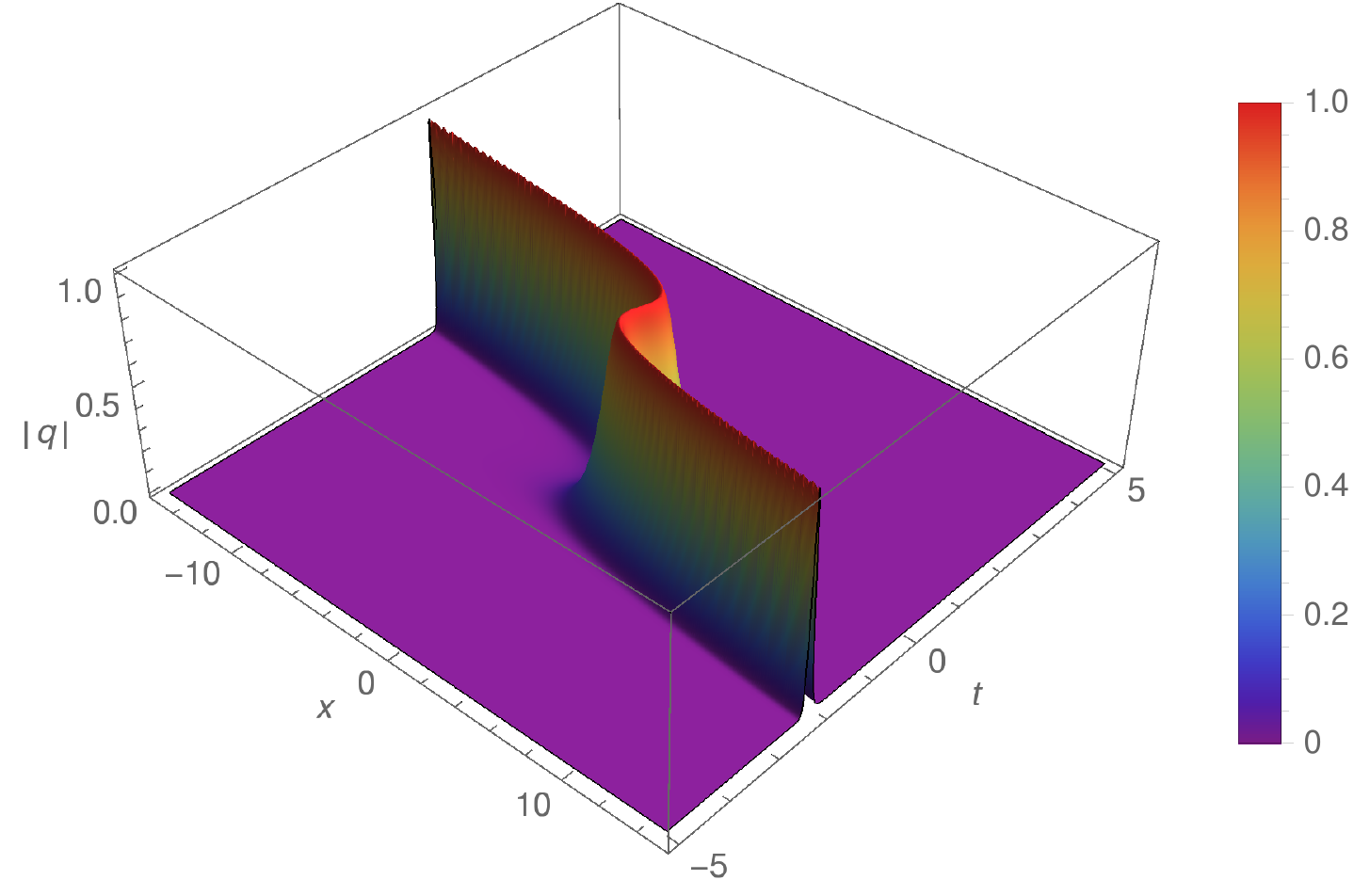}
\caption{}
\end{subfigure}%
\begin{subfigure}{.5 \textwidth}
\centering
\includegraphics[width=.8\textwidth]{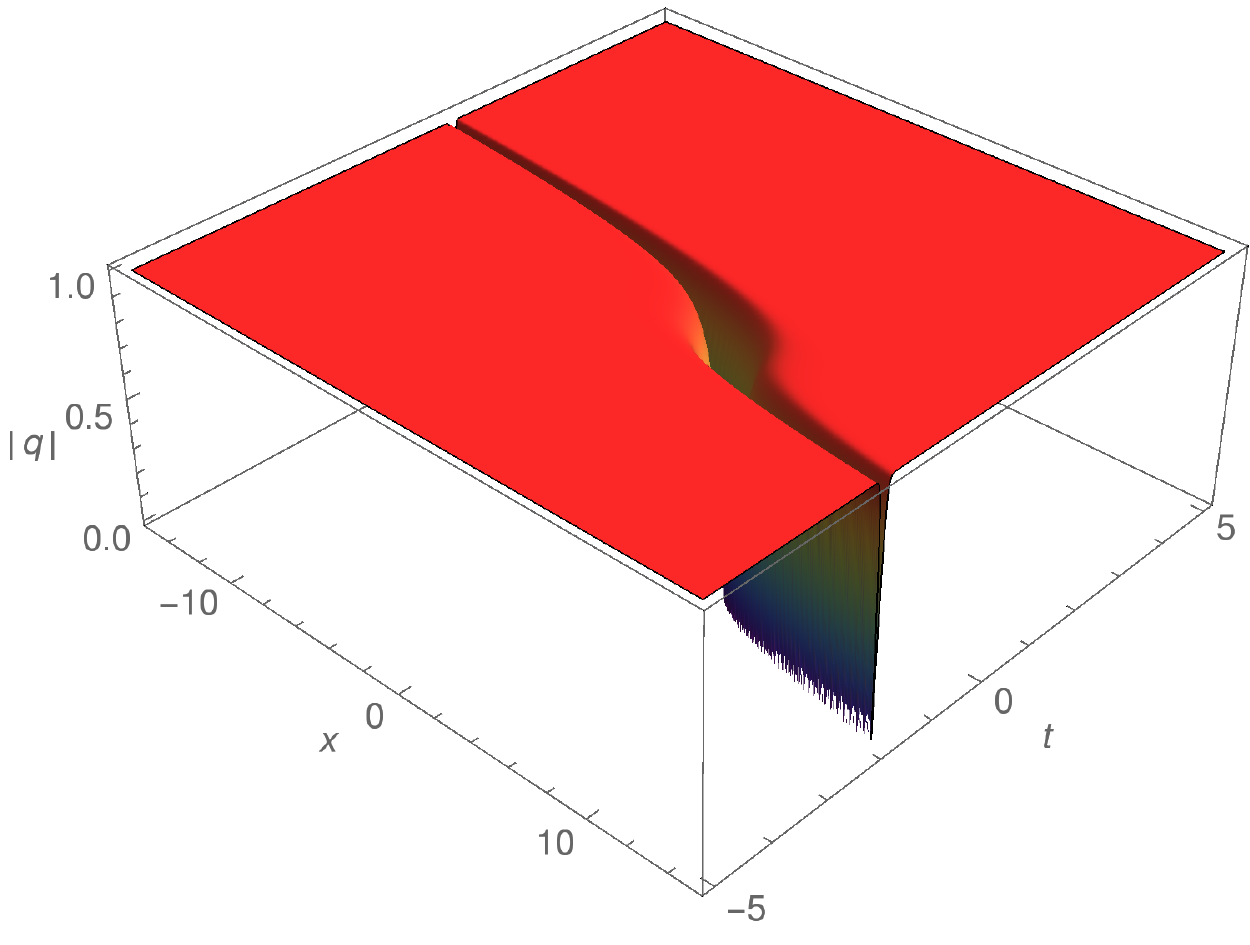}
\caption{}
\end{subfigure}
\caption{Evolution of a nonautonomous HNLSE bright(a) and dark(b) solitons with 
$D_{20}=R_{20}=1 $, $D_{30}=R_{30}=1 $, $\lambda = 1.2$ and soliton parameters $\eta=0.5$ and $\kappa= 0.9$ }
\label{fig-R3}
\end{figure*}

\begin{figure*}[htbp]
\begin{subfigure}{.5 \textwidth}
\centering
\includegraphics[width=1\textwidth]{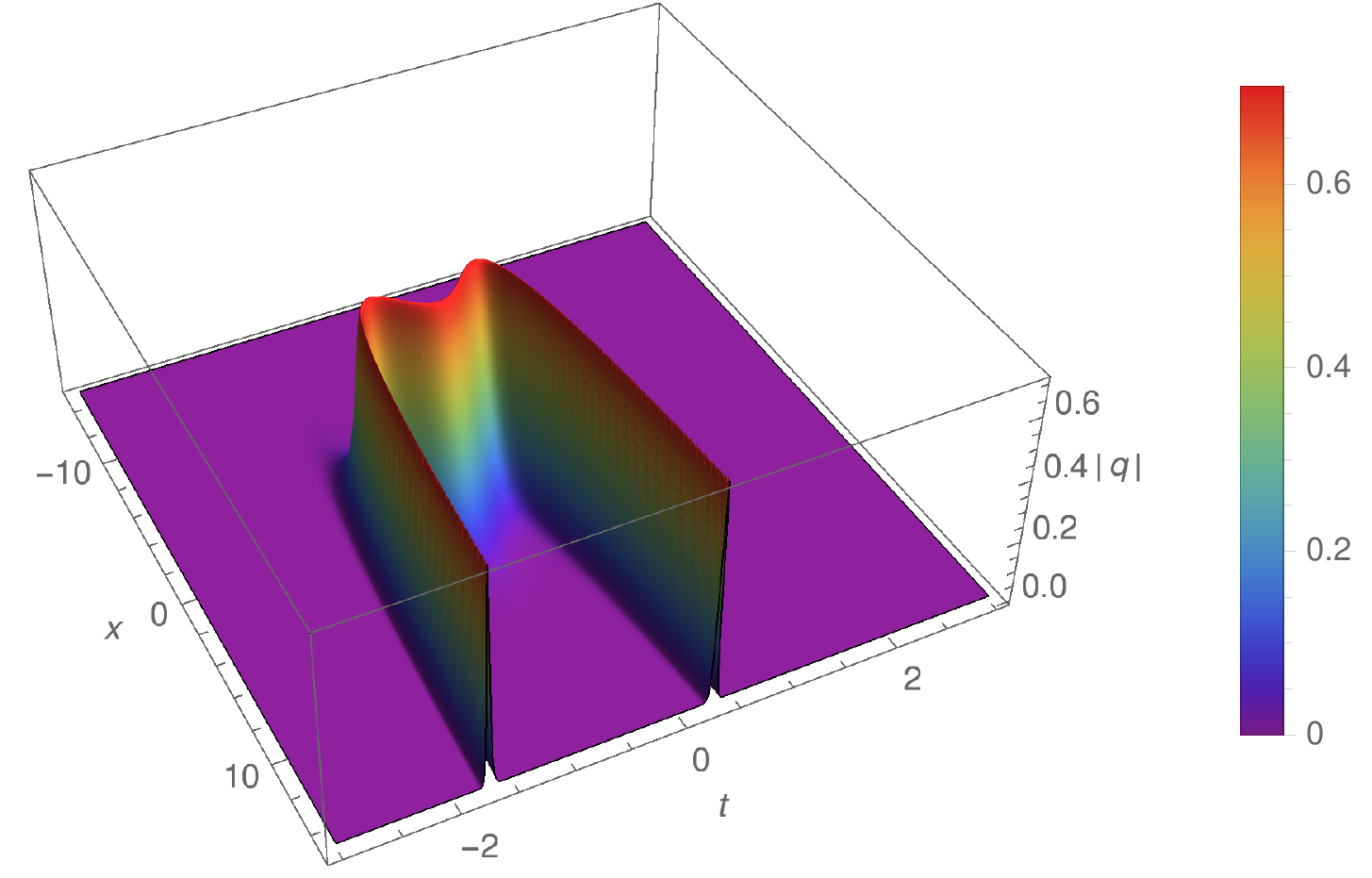}
\caption{}
\end{subfigure}%
\begin{subfigure}{.5 \textwidth}
\centering
\includegraphics[width=.8\textwidth]{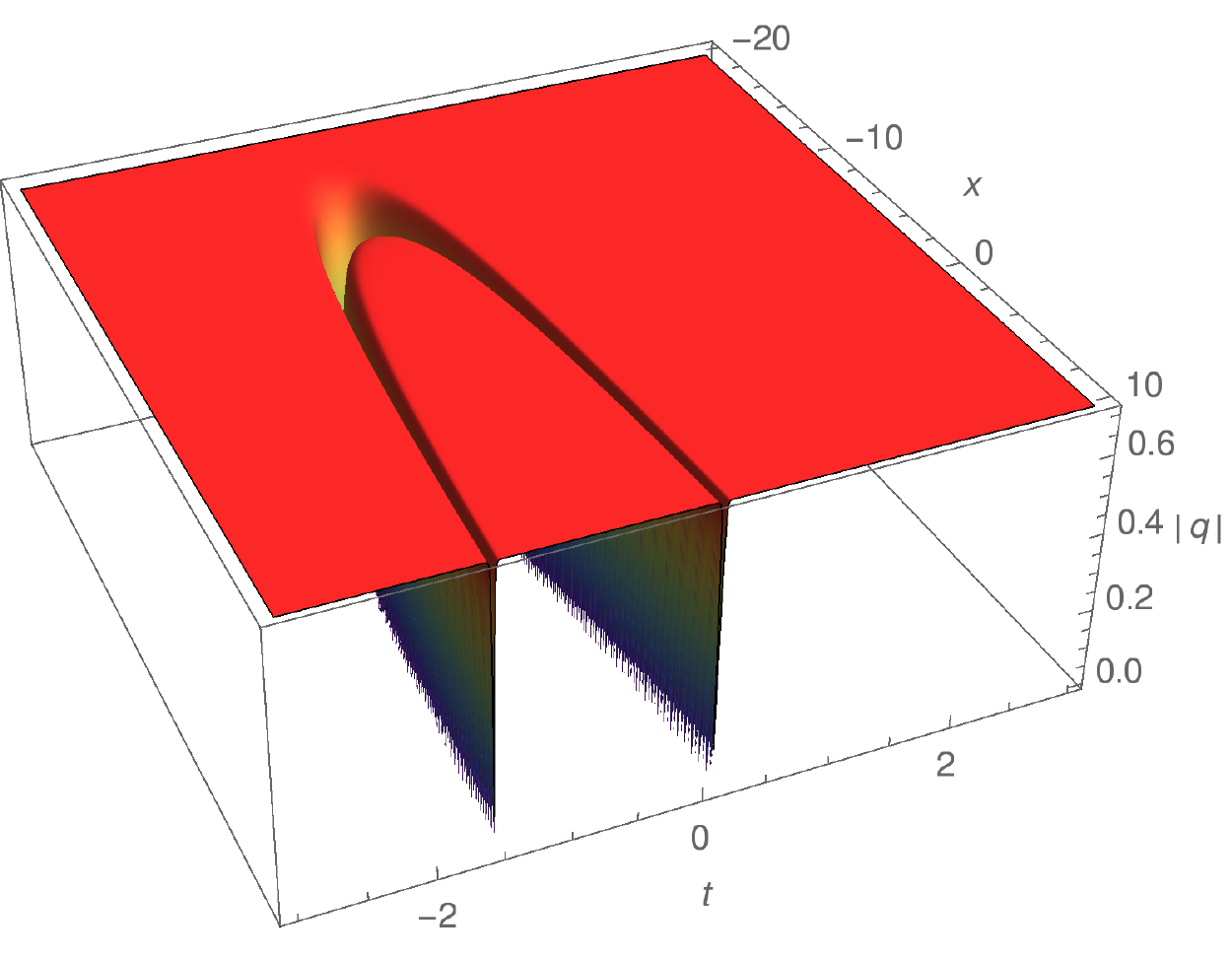}
\caption{}
\end{subfigure}
\caption{Evolution of a nonautonomous FLPDE bright(a) and dark(b) solitons with 
$D_{20}=3$, $R_{20}=6 $, $D_{30}=1$, $R_{30}=2 $, $D_{40}=2$, $R_{40}=4 $, 
$\lambda = 1.2$ and soliton parameters $\eta=0.5$ and $\kappa= 0.9$ }
\label{fig-R4}
\end{figure*}
Figures \ref{fig-R1} shows the evolution of  typical constant coefficient NLSE autonomous 
bright and dark soliton having a constant velocity.  Velocity of solitons of rest of the 
equations in the hierarchy  also remain constant with time as the soliton evolves.  
Figures (\ref{fig-R2} - \ref{fig-R4}) present   evolution 
of a bright soliton (a) and a dark  soliton (b) of first three equations of the NLSE hierarchy 
in presence of a linear potential. In the plot  absolute value  of the soliton's amplitude are plotted, 
suppressing the phase part. Notice that in presence of the potential,   NLSE  nonautonomous  
solitons are  "accelerated" (Fig. \ref{fig-R2}), that is the soliton group velocity 
is a function of $t$,  \ $v_g = 2 D_{20} \kappa + D_{20} \lambda t $. In this case  the GVD
parameter along  with the linear potential term 
account for the acceleration. On the other hand  nonautonomous HNLSE solitons are subjected 
to "accelerations" proportional to $t^2$ and $t^3$,  appearing in the argument of 
envelope functions $sech$  and $tanh$. Moving one up in the hierarchy, the figure shows that 
the nonautonomous  solitons  
are subjected to "accelerations"  proportional to $t^2$, $t^3$ and $t^4$. Thus the argument 
of envelope functions turn out to be a polynomial in $t$, the degree of the which increases
with the increase in the equation  order in the hierarchy. The "accelerations" also become a 
polynomial in $t$. and its degree increases as the order of the hierarchy. Note that the group 
velocity of nonautonomous  soliton of even  order NLSE hierarchy are fundamentally different from 
the  nonautonomous  soliton of odd  order NLSE hierarchy, that is in the former case 
soliton's velocity  is reversed after reaching the minima  and in the later case  
soliton's velocity increases in the same direction after reaching the minima. The number of 
minima increases with the order of hierarchy, see Figures (\ref{fig-R2} - \ref{fig-R4}). The 
inhomogeneous term in the NLSE hierarchy makes frequency, phase velocity also dependent 
on the time.
  
\section{Conclusion}
We propose a generalized reversible transformation between the  the generalized 
NLSE hierarchy and the  generalized forced NLSE hierarchy with  a linear 
potential term.  The reversible transformations allow to encompass 
inhomogeneous NLS, HNLS and higher order equations belonging to the class of 
nonisospectral family of inverse scattering problems into the  isospectral NLS class of 
equations and study them under the general mathematical framework. 
The corresponding extended concept of classical dark 
and bright solitons, accelerating in the linear potential of the forced hierarchy 
are obtained directly without resolving the nonisospectral  inverse scattering problem.
The transformation is not restricted to the case of dark and bright soliton only and 
may also be extended  to breather and other types of solitons.     
Soliton properties of the forced higher order NLSE family of equations are changed 
and the parameters such as  phase velocity, group velocity, frequency change 
with propagation and depends on time polynomially. 
We introduce a set of constraints for the  FLPDE  which are preserved 
after the transformations. Similar constraints are expected as one consider  higher order
equations. It has been noticed that the  the odd order inhomogeneous NLSE hierarchy 
are  fundamentally different from the even order inhomogeneous NLSE hierarchy. 
It would be interesting to extend the analysis to the coupled systems, that is 
Manakov equation hierarchy, which would be our future investigation.  We hope that our 
analysis open up the possibility of new applications  in physics.\\

%\bibliography{/home/sn/01-REVERSIBLE/refer}
%\begin{thebibliography99}

\end{document}